\documentclass[showpacs,twocolumn,pre]{revtex4}
\usepackage{psfrag,epsfig,amsfonts,amssymb,amsmath,graphicx,slashbox}
\usepackage{dcolumn}

\newcommand{\hr}{{\cal H}}
\newcommand{\ord}{{\cal O}}
\newcommand{\tr}{\mbox{tr}}

\begin{document}

\title{Typicality for generalized microcanonical ensembles}

\author{Peter Reimann}
\affiliation{Universit\"at Bielefeld, Fakult\"at f\"ur Physik, 33615 Bielefeld, Germany}

\begin{abstract}
For a macroscopic, isolated quantum system 
in an unknown pure state, the expectation 
value of any given observable is shown
to hardly deviate from the ensemble
average with extremely high probability
under generic equilibrium and non-equilibrium 
conditions.
Special care is devoted to the uncontrollable 
microscopic details of the system state.
For a subsystem weakly coupled to a large heat bath, 
the canonical ensemble is recovered
under much more general and realistic 
assumptions than those implicit in the usual
microcanonical description of the composite 
system at equilibrium.
\end{abstract}


\maketitle

Basic questions of statistical physics have gained renewed
interest with the discovery of various work and fluctuation 
theorems \cite{ref1}.
A further topic which has attracted much attention
concerns the foundation of the canonical formalism \cite{gol,pop,gemmer,early}.
One of its key ingredients consist in shifting the focus from
the traditional statistical equilibrium ensembles back to the
role and predictability of one single experimental realization 
of a system (and its environment), described theoretically by a 
quantum mechanical pure state.
In essence, the main message of the seminal works \cite{gol,pop},
for which the name ``canonical typicality'' has been coined in \cite{gol},
is as follows. Consider the usual canonical setup, i.e. an isolated 
``super-system'', compound of a ``large'' thermal bath B,
a comparatively ``small'' subsystem of actual interest S, and
a negligibly weak coupling between them.
Those energy eigenstates of the compound S+B with
energy-eigenvalues in the interval $[E-\Delta E, \, E]$
span a Hilbert space, from which we pick at random a 
pure state $|\psi\rangle$.
The corresponding projector $|\psi\rangle\langle\psi|$
gives rise to a mixed state $\rho_S$ of the subsystem S
by tracing out the bath degrees of freedom.
Now, the remarkable finding of Refs. \cite{gol,pop} is that
$\rho_S$ will be extremely close to the standard canonical 
density operator $\rho_{can}$ of the subsystem S for the 
overwhelming majority of random pure states $|\psi\rangle$, 
hence the name ``canonical typicality''.
In other words, whatever is the (unknown) pure state 
of the compound S+B, the outcome of any experiment
on the subsystem S is practically the same as if it 
were in the canonically mixed state $\rho_{can}$.
For a more detailed, precise, and also more general 
exposition we refer to the original Refs. \cite{gol,pop}.
Further, it should be pointed out that
essentially the same conclusion could be inferred from 
formula (C.17) of the formidable prior work \cite{gemmer}.
For less general system classes and/or 
after performing an additional time average,
closely related results have been obtained 
even earlier in Refs. \cite{early}.

Here, we will show that a quite similar ``typicality''
property already holds for any isolated system, even when it 
cannot be decomposed into subsystem S and bath B.
In the special case that such a decomposition is possible, 
the original ``canonical typicality'' is recovered by ``tracing out the bath'',
thereby shedding also new light on the role of entanglement.
A further main point is to abandon the quite 
unrealistic assumption of the previous works
\cite{gol,pop} that all energy eigenstates
belonging to the preset energy interval 
$[E-\Delta E,\, E]$ contribute to the pure state 
$|\psi\rangle$ with equal probabilities
and all the other energy eigenstates are excluded.
Rather, in reality one only knows the
occupation probabilities of the energy
levels very roughly and hence the unknown details 
should not matter in the final results.
This problem (and our solution) is clearly 
not restricted to the issue of typicality but 
concern the standard microcanonical formalism 
in general.

{\em Setup:}
We consider a quantum mechanical system, whose Hilbert 
space $\hr$ is spanned by the orthonormal basis 
$\{|n\rangle\}_{n=1}^N$, $N\leq\infty$.
Hence, any pure state $|\psi\rangle$ is of the form
\begin{equation}
|\psi\rangle = \sum \frac{c_n}{||c||}\, |n\rangle \ ,
\label{1}
\end{equation}
where $c_n\in {\mathbb C}$, 
$c:=(c_1,...,c_N)$, 
$||c||:=\sqrt{\sum |c_n|^2}$,
and the sum runs from $n=1$ to $N$.
The division by $||c||$ will be particularly 
convenient for our purposes.

As in Refs. \cite{gol,pop,gemmer,early}, we assume that the 
system is in some pure state $|\psi\rangle  \in\hr$ 
but we do not know which one.
In other words, the $c_n$ in (\ref{1}) are randomly 
drawn from some probability density $p(c)$.
Denoting the corresponding ensemble
average of any function $g(c)$ by
\begin{eqnarray}
\overline{g(c)}:=\int g(c)\, p(c)\, \prod_{n=1}^N d(\mbox{Re} c_n) \, d(\mbox{Im} c_n) \ ,
\label{1a}
\end{eqnarray}
the expectation value of an arbitrary observable 
$A=A^\dagger\, : \hr\to\hr$ takes the form
\begin{eqnarray}
\langle A\rangle & := & \overline{\langle \psi|A|\psi\rangle}
=\tr (\rho A)
\label{2}
\\
\rho & := & \overline{|\psi\rangle\langle\psi|} =\sum_{n,m} 
\overline{(c_nc_m^\ast/||c||^2)}\ |n\rangle\langle m| \ .
\label{3}
\end{eqnarray}
For infinite dimensional systems, well defined limits $N\to\infty$
in (\ref{1}-\ref{3}) and later on are tacitly taken for granted.

Next we turn to our two key assumptions regarding $p(c)$:
(i) The $c_n$ in (\ref{1}) are statistically independent
and, moreover, $c_n$ and $e^{i\varphi_n}c_n$ are equally likely
for arbitrary phases $\varphi_n$, or equivalently, $p(c)$ is of the form
\begin{equation}
p(c)=\prod_{n=1}^N p_n(|c_n|) \ .
\label{4}
\end{equation}
As a consequence, the density operator 
(\ref{3}) takes the form
\begin{eqnarray}
\rho = \sum \rho_n \, |n\rangle\langle n| \ , \ \rho_n := \overline{|c_n|^2/||c||^2}
\label{5}
\end{eqnarray}
with $\rho_n\geq 0$ and $\sum \rho_n =1$.
(ii) The mixed state $\rho$ has a low purity, i.e.
\begin{equation}
\tr  \rho^2 = \sum \rho_n^2 \ll 1 \ ,
\label{6}
\end{equation}
or equivalently, $\max\rho_n\ll 1$,
or equivalently, there are not just a few dominating 
$\rho_n$ (summing up to almost unity).
Before justifying these two assumptions, 
we show what can be concluded from them.

{\em Typicality:}
Our first objective is to show that $\langle\psi|A|\psi\rangle$ 
is typically very close to the average (\ref{2}), i.e.
\begin{equation}
\sigma_{\!\! A}^2 :=
\overline{[\langle\psi|A|\psi\rangle - \langle A \rangle]^2} 
\label{7}
\end{equation}
is small in an appropriate sense.
Postponing the formal proof to a later paper,
we adopt here a more heuristic line of reasoning,
similarly in spirit to Ref. \cite{gol}.
We first consider the deviation
of $s(c):=||c||^2$
from its average $\bar s = \overline{||c||^2}=\sum \overline{|c_n^2|}$.
Eq. (\ref{4}) implies that the $|c_n|^2$
are independent random variables and Eq. (\ref{6})
that a large number of them significantly contributes 
to the sum $s$.
Taking for granted finite variances
\begin{equation}
q_n:=\overline{(|c_n^2| - \overline{|c_n^2|})^2} / \, \overline{|c_n^2|}^2
=\overline{|c_n^4|} / \, \overline{|c_n^2|}^2 -1 \ ,
\label{8}
\end{equation}
the law of large numbers implies that
$s(c)-\bar s$ is an unbiased random variable 
with a very small standard deviation 
compared to $\bar s$.
Hence, for our present purpose of estimating
(\ref{7}) we can replace
$||c||$ in (\ref{1}) in very good 
approximation by $\bar s^{1/2}$.
As a consequence, one finds from (\ref{5}) that
$\overline{|c_n^2|} = \rho_n\, \bar s$.
Similarly, introducing (\ref{1}) with 
$||c||\simeq \bar s^{1/2}$ into (\ref{7}) and 
exploiting (\ref{4}) yields
\begin{eqnarray}
\sigma_{\!\! A}^2 & = & 
\sum_{n,m} \overline{|c_n^2\, c_m^2|}
\frac{(1-\delta_{nm})|\tilde A_{mn}|^2+\tilde A_{nn}\tilde A_{mm}}{\bar s^2} 
\nonumber
\\
\tilde A & := & A-\langle A\rangle  \ , \  \tilde A_{nm}:=\langle n|\tilde A|m\rangle \ .
\label{10}
\end{eqnarray}
Observing that $\overline{|c_n^2\, c_m^2|} =\rho_n\, \rho_m\, \bar s^2$
for $n\not=m$ according to (\ref{4}) and $\overline{|c_n^2|} = \rho_n\, \bar s$,
that $\overline{|c_n^4|}=(q_n+1)\, \rho_n^2 \bar s^2$ according to (\ref{8}),
and that 
$\sum \rho_n \rho_m\, \tilde A_{nn}\tilde A_{mm}=(\sum \rho_n\, \tilde A_{nn})^2=0$ 
according to (\ref{2},\ref{5},\ref{10}), yields
\begin{eqnarray}
\sigma_{\!\! A}^2 & = & 
\sum_{n,m} \rho_n \rho_m\, \tilde A_{nm}\tilde A_{mn} 
+ \sum (q_n-1)\, \rho_n^2 \, \tilde A_{nn}^2 \ .
\label{11}
\end{eqnarray}
Exploiting again (\ref{2},\ref{5},\ref{10}),
the first sum can be identified with $\tr  (\rho\tilde A)^2$. 
Performing this trace with the help of the eigenvectors $|\nu\rangle$ 
and eigenvalues $\tilde a_\nu$ of $\tilde A$ yields 
$\sum \tilde a_\nu \tilde a_\mu\, 
|\langle\nu|\rho|\mu\rangle|^2$.
From (\ref{10}) it follows that
$\tilde a_{min}:=\min\tilde a_\nu \leq 0$, 
$\tilde a_{max}:=\max\tilde a_\nu \geq 0$, 
and hence $\Delta_{\! A}:=\tilde a_{max}-\tilde a_{min}\geq |\tilde a_\nu|$
for any $\nu$.
As a consequence, the first sum in (\ref{11}) is 
bounded by $\Delta_{\!\! A}^2 \, \tr \rho^2$.
Similarly, in the second sum we exploit that 
$\tilde A_{nn}^2\leq \Delta_{\! A}^2$,
yielding
\begin{eqnarray}
\sigma_{\!\! A}^2 & \leq & \Delta_{\! A}^2 \,  (\max_n q_n)\,  (\tr  \rho^2) \ .
\label{12}
\end{eqnarray}
In the special case that $\rho$ is of the standard 
microcanonical form (see below), the same result also 
follows from Eq. (C.17) in \cite{gemmer}.

Chebyshev's inequality implies for any given $\epsilon>0$ 
that $\sigma_{\!\! A}^2/\epsilon^2$ is an upper bound for the 
probability that
$|\langle\psi|A|\psi\rangle-\langle A\rangle|$
exceeds $\epsilon$.
Exploiting (\ref{12}), one finally infers for $K$ 
observables $\{A_k\}_{k=1}^K$ and any $\epsilon>0$ that 
\begin{eqnarray}
& & \mbox{Prob}\left(\max_{k\leq K}\left|\langle\psi|A_k|\psi\rangle-\langle A_k\rangle\right|/
\Delta_{\! A_k}
\geq \epsilon \right)
\nonumber
\\
& & 
\leq K\,  (\max_n q_n)\,  (\tr  \rho^2)/\epsilon \ .
\label{12a}
\end{eqnarray}
This is the first main result of our paper.
With (\ref{10}) one sees that $\Delta_{\! A}$ equals
the difference between the maximal and minimal eigenvalues 
$A$ and hence quantifies the full range of all a priori 
possible values of $\langle\psi|A|\psi\rangle$.
In (\ref{12a}) we tacitly excluded trivial observables
$A_k$ with $\Delta_{\! A_k}=0$.
The $q_n$ in (\ref{8}) and hence $\max q_n$ are dimensionless,
non-negative numbers, typically of the order of unity. 
E.g. any Gaussian factor $p_n$ in (\ref{4}) yields 
$q_n=1$.
Hence, (\ref{12a}) with (\ref{6}) imply typicality:
a randomly sampled pure state $|\psi\rangle\in\hr$
is very likely to yield expectation values 
$\langle\psi|A_k|\psi\rangle$ very close to the ensemble averages 
$\langle A_k\rangle=\tr(\rho A_k)$ simultaneously for a quite large 
number $K$ of arbitrary but fixed observables $A_1,...,A_K$.

{\em Generalized microcanonical formalism:}
So far, typicality (\ref{12a}) applies to any quantum mechanical
system satisfying (\ref{4}) and (\ref{6}).
Next we specifically justify these assumptions (\ref{4}), (\ref{6})
for an isolated system at thermal equilibrium  
with $f=\ord(10^{23})$ degrees of freedom and with $|n\rangle$ 
being the eigenvalues of the Hamiltonian 
$H=\sum E_n\,|n\rangle\langle n|$, $E_n\geq E_{n-1}$, $E_1>-\infty$.

The assumption that coefficients $c_n$ and 
$e^{i\varphi_n}c_n$ occur with equal 
probability in (\ref{1}) is quite suggestive.
Indeed, upon time evolution, the eigenvectors 
$|n\rangle$ acquire factors of the form 
$e^{-iE_n t/\hbar}$. Taking for granted 
that $p(c)$ does not change with time at
thermal equilibrium, the invariance 
under $c_n\mapsto e^{i\varphi_n}c_n$ 
follows under rather mild and 
generic incommensurability conditions 
for the $E_n$. 
Exploiting this property, one
readily concludes that $\bar c_n=0$ for 
all $n$ and that $\overline{c_n c_m}=0$ 
for all $n\not=m$. 
In other words, the $c_n$ are uncorrelated. 
This does not yet imply independence in principle, 
but in practice it almost always does, 
and hence assumption (\ref{4}) is reasonable.

The starting point of the seminal works \cite{gol,pop} 
is the assumption that in (\ref{1}) all coefficients 
$c_n$ corresponding to energies $E_n$ within some 
preset energy interval $[E-\Delta E,\, E]$ are
``equally likely'', while all other $c_n$ are zero.
Denoting by $c'$ the vector of all non-zero $c_n$
and by $c''$ those which must be zero, this means that
$p(c)$ can be so chosen that
all $c'$ of equal length $||c'||$ must be equally probable
and thus $p(c)$ must be of the form $g(||c'||)\delta(c'')$
for some (properly normalized, non-negative) function $g$.
Further, the division by $||c||$ in (\ref{1}) implies that 
any such $g$ actually yields the same distribution of 
vectors $|\psi\rangle$.
Choosing a Gaussian $g$, the $c_n$ can thus 
without loss of generality be considered \cite{gol} as
independent, Gaussian, and satisfying (\ref{4}).
Moreover, $\rho$ from (\ref{5}) becomes the standard 
microcanonical density operator $\rho_{mic}$ with equal
weights $\rho_n>0$ if $E_n \in [E-\Delta E,\, E]$ 
and $\rho_n=0$ otherwise \cite{diu}.

On one hand, our present approach thus includes
the standard microcanonical formalism \cite{diu} 
and the starting point of Refs. \cite{gol,pop} 
as special cases.
On the other hand, the above observation that an entire class 
of different $p(c)$ actually yields 
-- due to the division by $||c||$ in (\ref{1})  -- 
the same distribution of vectors $|\psi\rangle$ still remains true, 
and hence the assumption that one of those equivalent $p(c)$ 
satisfies (\ref{4}) is very weak indeed.
On top of that, our above proof of typicality can even
be significantly generalized beyond the independence
assumption (\ref{4}) itself.
In particular, the Gaussian adjusted projected measures 
(GAP) from \cite{gol06} are still admissible.

Before turning to condition (\ref{6}), 
we recall some standard notions and general properties 
regarding the energy spectrum $\{E_n\}$ \cite{diu}.
Denoting the number of states within
$[E-\Delta E,\, E]$ by $\Omega(E)$ and
Boltzmann's constant by $k_B$, 
entropy and temperature follow as
\begin{equation}
S(E):=k_B \ln \Omega (E)\ , \ T(E):=1/S'(E) \ .
\label{13}
\end{equation}
One finds \cite{diu} that $\Omega(E)$
is a very rapidly increasing function of $E$
with typical values in range of 
$10^{\ord(f)}$,  $f=\ord(10^{23})$.
Hence, $\Omega(E)$ is largely independent of 
$\Delta E$ (provided $\Delta E\gg k_BT$),
and its derivative can be identified with
the density of states,
\begin{equation}
\Omega' (E) = \sum\delta(E_n-E) \ , 
\label{14}
\end{equation}
where the delta functions are slightly washed 
out to yield smooth functions in (\ref{13}).

In view of this tremendous density of energy 
levels $E_n$, it is indeed quite convincing 
that no real system can be prepared such that
just a few of them are populated with
appreciable probability $\rho_n$, 
implying that (\ref{6}) is indeed 
satisfied.
In fact, an even stronger statement is quite 
plausible and will be derived dynamically elsewhere,
namely that these populations $\rho_n$ 
can be written in the form
\begin{equation}
\rho_n=h(E_n)
\label{15}
\end{equation}
with a smooth function $h$, exhibiting
appreciable variations only on scales 
much larger that $E_n-E_{n-1}$.

Due to (\ref{15}), the energy distribution
$p(E):=\langle \delta(H-E)\rangle$
can be rewritten with (\ref{2},\ref{5})
as $\sum \rho_n \, \delta(E_n-E)=h(E) \sum\delta(E_n-E)$
and, after washing out the delta functions
according to (\ref{14}), as
\begin{equation}
p(E)=h(E)\, \Omega'(E) \ .
\label{16}
\end{equation}
In reality, after the experimentalist has prepared
the system as carefully as possible, the only available 
knowledge about $h$ and hence $\rho$ is that the 
probability density (\ref{16}) exhibits a relatively 
sharp peak (but still very wide compared to $E_n-E_{n-1}$).
All further details of $p(E)$ are 
completely fixed by the given experimental 
setup, but it is impossible to know them.
The only way out is to verify that these 
details ``do not matter''.
Experimentally, this seems indeed to be the case,
but theoretically it has apparently not been 
demonstrated so far.
On the contrary, for the usually considered 
$\rho_{mic}$, the concomitant details of $p(E)$ are
in fact quite unrealistic.

As a first example, we show that the celebrated
relation
\begin{equation}
-k_B\tr(\rho\ln\rho) = S(E^\ast)
\label{17}
\end{equation}
indeed holds for any sharply peaked $p(E)$ 
with $E^\ast$ located in the peak region:
Exploiting (\ref{5}) and (\ref{15}) yields
\begin{eqnarray}
& & \tr(\rho\ln\rho)=\sum h(E_n) \ln h(E_n) =
\nonumber
\\
& & 
= \int dE\, h(E)\ln h(E)\, \sum\delta(E_n-E) \ .
\nonumber
\end{eqnarray}
Due to (\ref{14}) and (\ref{16}) we can conclude that
\begin{eqnarray}
\tr(\rho\ln\rho)=\int dE\, p(E)\ln h(E) \ .
\label{18}
\end{eqnarray}
The integrand is dominated by the sharp peak of $p(E)$
since the possibly comparable variations of $h(E)$ 
(cf. (\ref{16})) are tamed by the logarithm.
Hence, there exists an energy $E^\ast$ within the
peak region with the property
\begin{eqnarray}
\int dE\, p(E)\ln h(E) = \ln h(E^\ast)\int dE\, p(E)\ .
\nonumber
\end{eqnarray}
Likewise, Eq. (\ref{16}) and the normalization of $p(E)$
imply
\begin{eqnarray}
1= \int dE\, p(E) = p(E^\ast)\,  \epsilon = h(E^\ast)\, \Omega'(E^\ast)\, \epsilon \ ,
\label{19}
\end{eqnarray}
where $\epsilon$ essentially represents the peak width of $p(E)$.
Eqs. (\ref{13}) imply $\Omega'=\Omega/k_BT$, yielding with (\ref{18})-(\ref{19})
\begin{eqnarray}
\tr(\rho\ln\rho)= -\ln \Omega(E^\ast) - \ln(\epsilon/k_BT(E^\ast)) \ .
\label{20}
\end{eqnarray}
Since $\ln \Omega=\ord(10^{23})$
(see below (\ref{13})), the last term in (\ref{20}) is negligible 
for any realistic $\epsilon$ and with (\ref{13})
our proof of (\ref{17}) is completed.
A somewhat similar calculation has been performed in chapter 12.3.2 of 
\cite{gemmer} but its purpose and physical content is quite different.

An analogous line of reasoning yields
$\tr\rho^2\approx1/\Omega (E^\ast)=10^{-\ord(10^{23})}$,
i.e. typicality (\ref{12a}) is extremely well satisfied
for a very large number $K$ of observables.

{\em Canonical formalism:}
As in the introduction, we consider a subsystem S
in weak contact with a much larger bath B, resulting 
in a compound S+B with a product Hilbert space
$\hr = \hr_S \otimes \hr_B$ and a
Hamiltonian $H=H_S\otimes 1_B + 1_S\otimes H_B$,
where $1_S$ is the identity on 
$\hr_S$ and similarly for $1_B$.
Given $H_S=\sum E_j^S\, |j\rangle_{S\, S}\langle j|$ and
$H_B=\sum E_k^B \, |k\rangle_{B\, B}\langle k|$ 
we thus have $H=\sum_{jk} E_{jk} |jk\rangle \langle jk|$
with $E_{jk}:=E_j^S+E_k^B$ and
$|jk\rangle:=|j\rangle_S|k\rangle_B$,
i.e. previous labels $n$ now become $jk$.
Since only subsystem observables 
$A=A_S\otimes 1_B$ are of interest,
(\ref{2}) can be rewritten as
\begin{eqnarray}
\langle A\rangle = \tr_S(\rho_{can}\, A_S) \ , \ 
\rho_{can}:= \tr_B(\rho) \ ,
\label{21}
\end{eqnarray}
with $\tr_S$ the partial trace over $\hr_S$, and similarly for $\tr_B$.
Likewise, 
$\langle\psi|A|\psi\rangle = \tr (|\psi\rangle\langle\psi|\, A)$
can be rewritten as
\begin{eqnarray}
\langle\psi|A|\psi\rangle  = \tr_S(\rho_S A_S) \ , \
\rho_S  :=  \tr_B (|\psi\rangle\langle\psi|) \ .
\label{22}
\end{eqnarray}
Eq. (\ref{12a}) implies ``canonical typicality'' 
in the sense that for the vast majority of pure states $|\psi\rangle$
of the compound S+B, the corresponding mixed
state of the subsystem $\rho_S$ yields practically
the same result for $K$ subsystem observables 
$A_{S,1},...,A_{S,K}$ 
as the ensemble averaged mixed state 
$\rho_{can}$.
If the bath B is sufficiently much larger than
the subsystem S then the extremely low purity of 
the compound S+B implies typicality even for 
{\em all} possible subsystem observables $A_S$, 
giving rise to a natural metric \cite{pop}
according to which the reduced density operator 
$\rho_S$ itself is close to $\rho_{can}$.

Finally, one finds that any $p(E)$ in (\ref{16}) with a 
sharp peak near $E^\ast$ results in the same canonical 
density matrix
\begin{equation}
\rho_{can}=Z^{-1} \exp\{-H_S/k_BT(E^\ast)\} \ .
\label{23}
\end{equation}
The main line of reasoning to prove (\ref{23})
is analogous to (\ref{17})-(\ref{20}), while
the somewhat more tedious details will be
presented elsewhere.
In particular, (\ref{23}) implies that the expectation
value (\ref{21}) of arbitrary subsystem observables
$A_S$ are indeed independent of any further details 
of $p(E)$.

{\em Conclusions:}
We have shown that the overwhelming majority
of pure states yields practically identical 
expectation values for any given (not extremely large)
set of observables under conditions which are 
generically satisfied for isolated macroscopic systems at
thermal equilibrium.
A second main result is that the experimentally
uncontrollable and hence unknown microscopic
details of the system state are indeed 
irrelevant.
In particular, for the practically most important
system-plus-bath setup, 
the canonical ensemble (\ref{23}) is recovered
under much more general and realistic 
assumptions than those implicit in the usual
microcanonical description of the composite 
system.

The finding that (\ref{17}) does not depend 
on the unknown details of the equilibrium ensemble 
$\rho$ also sheds new light on 
the usual information theoretical ``derivation'' 
of the microcanonical ensemble \cite{diu}: 
While $\rho_{mic}$ indeed minimizes the information 
functional $\tr\rho\ln\rho$, the information
content of many other $\rho$'s is almost
equally low and one cannot conclude that 
practically only the exact minimum $\rho_{mic}$
occurs in reality.

The present approach generalizes the seminal
works \cite{gol,pop} on canonical typicality 
in two respects:
The system needs not be a subsystem-plus-bath
compound, and the equilibrium ensemble need not
be of the quite particular microcanonical form.

Given a compound S+B in a pure state $|\psi\rangle$,
a well know consequence of entanglement between 
subsystem S and bath B is a mixed state $\rho_S$ 
after tracing out the bath according to (\ref{22}).
While entanglement has been proposed as main 
origin of canonical typicality in Refs. \cite{pop},
our present findings suggest that the main root
is the typicality property of the entire compound,
which is in turn not entangled with any further 
system.

We close with a simple but quite interesting observation
regarding systems {\em out of equilibrium}:
Specifically, assume that the system is isolated and 
has reached equilibrium for times $t\leq 0$, while for
$t>0$ an external perturbation sets in \cite{ref1}, 
giving rise to an explicitly time dependent Hamiltonian $H(t)$ 
and a corresponding propagator $U(t_2,t_1)$.
Instead of propagating the states $|\psi\rangle$ beyond
$t=0$, we can switch to the Heisenberg picture and
instead propagate the observables. In this way,
by replacing for any given $t>0$ the original 
observable $A$ by $U^\dagger(t,0)AU(t,0)$, all the
equilibrium typicality properties at $t=0$ immediately 
carry over to the out of equilibrium situation for 
$t>0$.
For not explicitly time dependent $A$, the spectrum 
remains invariant under time propagation, 
and hence (\ref{12a}) remains valid for any 
given $t>0$ with $\rho$ being 
the equilibrium density operator at $t=0$.

\end{document}